\documentclass{article}

\usepackage{arxiv}

\usepackage[utf8]{inputenc} 
\usepackage[T1]{fontenc}    
\usepackage{hyperref}       
\usepackage{url}            
\usepackage{booktabs}       
\usepackage{amsfonts}       
\usepackage{nicefrac}       
\usepackage{microtype}      
\usepackage{lipsum}		
\usepackage{graphicx}
\usepackage{natbib}
\usepackage{doi}
\usepackage[acronym]{glossaries}

\makeglossaries
\newacronym{pdf}{PDF}{probability density function}
\newacronym{cdf}{CDF}{cumulative distribution function}
\newacronym{ml}{ML}{machine learning}
\newacronym{nf}{NF}{Normalising Flow}
\newacronym{kl}{KL}{Kullback-Leibler}
\newacronym{js}{JS}{Jensen-Shannon}
\newacronym{ns}{NS}{Nested Sampling}

\title{Normalising Flows for Bayesian Gravity Inversion}

\author{ Henrietta Rakoczi\\
	School of Physics and Astronomy\\
	University of Glasgow\\
	Glasgow, UK\\
	\texttt{h.rakoczi.1@research.gla.ac.uk} \\
	\And
    Abhinav Prasad \\
	School of Physics and Astronomy\\
	University of Glasgow\\
	Glasgow, UK\\
 	\And
    Karl Toland \\
	School of Physics and Astronomy\\
	University of Glasgow\\
	Glasgow, UK\\
 	\And
    Christopher Messenger\\
	School of Physics and Astronomy\\
	University of Glasgow\\
	Glasgow, UK\\
 	\And
    Giles Hammond \\
	School of Physics and Astronomy\\
	University of Glasgow\\
	Glasgow, UK
}

\renewcommand{\shorttitle}{Normalising Flows for Bayesian Gravity Inversion}

\hypersetup{
pdftitle={A template for the arxiv style},
pdfsubject={q-bio.NC, q-bio.QM},
pdfauthor={David S.~Hippocampus, Elias D.~Striatum},
pdfkeywords={First keyword, Second keyword, More},
}

\begin{document}
\maketitle

\begin{abstract}
Gravity inversion is a commonly applied data analysis technique in the field of geophysics. While machine learning methods have previously been explored for the problem of gravity inversion, these are deterministic approaches returning a single solution deemed most appropriate by the algorithm. The method presented here takes a different approach, where gravity inversion is reformulated as a Bayesian parameter inference problem. Samples from the posterior probability distribution of source model parameters are obtained via the implementation of a generative neural network architecture known as Normalising Flows. Due to its probabilistic nature, this framework provides the user with a range of source parameters and uncertainties instead of a single solution, and is inherently robust against instrumental noise. The performance of the Normalising Flow is compared to that of an established Bayesian method called Nested Sampling. It is shown that the new method returns results with comparable accuracy 200 times faster than standard sampling methods, which makes Normalising Flows a suitable method for real-time inversion in the field. When applied to data sets with high dimensionality, standard sampling methods can become impractical due to long computation times. It is shown that inversion using Normalising Flows remains tractable even at 512 dimensions and once the network is trained, the results can be obtained in $O(10)$ seconds.
\end{abstract}


\printglossary[type=\acronymtype]

\section{Introduction}
Gravimetry is a method that measures differences in the gravitational acceleration on the Earth's surface to a very high precision. Measuring gravity can be a useful probe when the detection or characterisation of underground features is desired as spatial or temporal changes in gravity can indicate fluctuations in the density contrast between regions or objects of interest (\cite{Torge(1989)}). It can act as a complementary tool to other methods, such as magnetic field or seismic measurements (\cite{Milano(2021)}), and can offer an alternative when other methods are not viable. Gravimetry can be used for civil engineering purposes, such as the detection of underground man-made structures (\cite{Pearse(2021)}, \cite{Eppelbaum(2010)}), geophysical applications such as volcano monitoring (\cite{Cabone(2020)}), or oil and gas exploration~(\cite{Gadirov(2022)}). In addition, gravimetry can not only be used for detection and monitoring, but also for the deduction of the parameters of the underground density distribution.

Modelling underground structures from geophysical measurements is an inverse problem, where the computation of a mapping from the measured data to the model governing the data is desired. In the case of gravimetry, this can be formulated as an inverse source problem, such as
\begin{equation}
\label{eqn:sourceproblem}
\boldsymbol{\theta} = A^{-1}(\boldsymbol{d})
\end{equation}
where $\boldsymbol{\theta}$ is the source model or underground density distribution, $\boldsymbol{d}$ is the measured gravity data, and $A^{-1}$ is the inverse mapping (\cite{Zhdanov(2002)}). In gravimetry it is often necessary that the model is defined in higher dimensions than the data itself, which results in an ill-posed problem. When measuring the gravitational acceleration of a 3-dimensional model on a flat surface, some information about the source is inherently lost and the solution becomes non-unique. For instance, a massive source buried deep underground might result in a similar gravity profile as a much smaller, shallow object. These degeneracies cannot easily be broken, but constraints can be introduced based on prior knowledge about the characteristics of the source and local geology. Apart from this problem of uniqueness, another concern in gravity inversion is the stability of the solution in the presence of measurement noise. The solution is deemed stable if it does not vary significantly within the noise range of the survey.

Traditional linear inversion algorithms apply the least squares method to iteratively optimise the source model parameters until a solution is found (\cite{Sen(2013)}, \cite{Lines(1984)}). The stability and consistency of results from these methods are often susceptible to variations in the measurement data and require the introduction of regularisation functions (\cite{Tikhonov(1977)}), which encode prior information and need to be carefully chosen for the specific problem (\cite{Zhdanov(2009)}). While standard methods default to a single solution, we know that a family of solutions exist due to non-uniqueness and noise. Therefore, additional considerations are required to approximate uncertainties on the result.

Alternatively, gravity inversion can be reformulated as a Bayesian parameter inference problem. Within this framework, the parameters of the source model ($\boldsymbol{\theta}$) are treated as random variables and a \acrlong{pdf} (\acrshort{pdf}) of these parameters is inferred to describe the solution to the inversion. Due to its probabilistic nature, the method can naturally account for instrumental noise and maintain stability. Any previous knowledge of the source and characteristics of the site can also be encoded in the prior, which is the assumed \acrshort{pdf} of the parameters of interest before any knowledge is obtained from the measurements. In addition, using this method we obtain an ensemble of solutions in the form of samples from the Bayesian posterior, instead of a single instance. This allows the user to have a better understanding of the uncertainties associated with the problem, obtain statistical measures, and visualise a whole range of possible models that could explain the measured data. Bayes' Theorem (\cite{Bayes(1763)}) describes the posterior \acrshort{pdf}, $p(\boldsymbol{\theta}|\boldsymbol{d},A)$ as
\begin{equation}
\label{eqn:bayestheorem}
p(\boldsymbol{\theta}|\boldsymbol{d},A) = \frac{p(\boldsymbol{\theta}|A)p(\boldsymbol{d}|\boldsymbol{\theta},A)}{p(\boldsymbol{d}|A)}
\end{equation}
where $\boldsymbol{d}$ is the observed data, $A$ is the assumed forward model, $\pi(\boldsymbol{\theta}) \equiv p(\boldsymbol{\theta}|A)$ is the prior and $L(\boldsymbol{\theta}) \equiv p(\boldsymbol{d}|\boldsymbol{\theta},A)$ is the likelihood. The denominator is a normalising constant, called the Bayesian evidence. To obtain the evidence, the computation of the integral 
\begin{equation}
\label{eqn:bayesevidence}
Z = p(\boldsymbol{d}|A) = \int L(\boldsymbol{\theta}) \pi(\boldsymbol{\theta}) d\boldsymbol{\theta}
\end{equation}
over the whole parameter space is required, which can become intractable for high-dimensional spaces. Therefore, various methods have been developed to approximate this term or avoid computing it altogether.

Stochastic sampling methods (\cite{Sivia(2006)}) are commonly applied to approximate the posterior. Some of these methods have been successfully applied to gravity inversion (\cite{Brown(2015)}), however the computation time increases with the complexity and dimensionality of the posterior (\cite{Rodgers(2017)}). The application of \acrlong{ml} (\acrshort{ml}) can potentially improve on sampling methods by reducing computation time and allowing the approximation of posterior probability distributions even for high-dimensional data. In this work, a specific type of generative neural network model implementing \acrlong{nf}s (\acrshort{nf}) (\cite{Rezende(2016)}) is investigated for applications in gravity inversion. \cite{Valentine(2022)} identified variational inference using generative models as a field of rapid development and interest within the geophysics community.  While various \acrshort{ml} methods have been applied to deterministic, single-solution gravity inversion (\cite{Yang(2022)}, \cite{Zhang(2022)}), the application of neural networks for probabilistic gravity inversion is still in its early stages. More specifically, \acrshort{nf}s were recently explored for the inverse problem in seismic tomography (\cite{Zhao(2021)}), but until now these not have been tested for gravity data analysis.

\section{Methodology}\label{methodology}
In this section different aspects of the methodology used in this work are discussed. In Section \ref{nestedsampling}, the \acrlong{ns} (\acrshort{ns}) method is described, which is a well-established algorithm used for probabilistic modelling and Bayesian inference and is applied here as a benchmark algorithm. In Section \ref{normalisingflows}, the mathematical framework of the \acrshort{nf} method is presented. Finally, in Section \ref{forwardmodel}, the applied forward model is discussed. The principles of operation of the inversion algorithms are explained here, while the specifics of the implementation are outlined in Section \ref{implementation} of this paper.

\subsection{Nested Sampling}\label{nestedsampling}
In Bayesian analysis, evaluating the evidence and the posterior is trivial for low dimensional data, described by a small number of parameters and by simple probability distributions. For higher dimensions and more complex distributions, the direct computation of the evidence and the posterior can become intractable and other approaches are required. Algorithms such as Markov-Chain Monte Carlo (\cite{Hastings(1970)})
were designed to avoid the evaluation of the evidence when only samples from the posterior distribution are needed. Meanwhile, \acrshort{ns} (\cite{Skilling(2006)}), aims to efficiently compute an approximation of the evidence from previously defined analytical likelihood and prior functions. In addition, as is used in this work, samples from the posterior can be produced as a by-product of the evidence evaluation. The idea behind \acrshort{ns} is to reduce the multi-dimensional integral defining the evidence ($Z$) in Equation \ref{eqn:bayesevidence} to a one-dimensional problem by defining the {\it prior mass} ($X$) via $dX = \pi(\boldsymbol{\theta}) d\boldsymbol{\theta}$. The likelihood can then be expressed in terms of the prior mass instead of the model parameters, so that $L(X(\lambda)) \equiv \lambda $, where
\begin{equation}
\label{eqn:priormass}
X(\lambda) = \int_{L(\boldsymbol{\theta})>\lambda} \pi(\boldsymbol{\theta}) d\boldsymbol{\theta}
\end{equation}
is the prior mass enclosed within the region of the parameter space bounded by the likelihood contour $\lambda$. Therefore, Equation \ref{eqn:bayesevidence} becomes
\begin{equation}
\label{eqn:evidencechangeofvariable}
Z = \int_0^1 L(X) dX,
\end{equation}
which can be computed using simple numerical integration methods.

The \acrshort{ns} algorithm starts with sampling $N$ {\it live points} from the prior, and the associated likelihood is computed for of each of these points. The point with the lowest likelihood ($L_1$) is identified and removed from the set. A new sample is drawn to replace it, which is only accepted if its associated likelihood is $ \ge L_1$. Setting $L_1$ to be a threshold shrinks the data space volume that the samples are drawn from. This process is repeated in iterations, tightening the likelihood bound each time, until the stopping criteria is reached. For each iteration, the likelihood limit ($L_i$) as well as the changes in the prior mass contained within this limit ($\Delta X_i$) are stored. The integral in Equation \ref{eqn:evidencechangeofvariable} then can be approximated using the trapezoid rule, such as $Z = \sum L_i \Delta X_i$ (\cite{Buchner(2023)}). A useful by-product of the process is that the removed live points can be used to obtain posterior samples. At the end of the process, samples are drawn from the assembly of all previous and final points with the $i^{th}$ point weighted by the factor $L_i \Delta X_i$. These samples are representative of the posterior.

The algorithm requires the likelihood and prior functions associated with the problem to be pre-defined, which encode some previously known information about the noise associated with the gravity survey itself and the constraints that can be placed on the source parameters before any knowledge is gained from the measured data. Sampling methods have been successfully applied to gravity (\cite{Brown(2015)}, \cite{Rodgers(2017)}) and combined gravity and magnetic (\cite{Ghalenoei(2021)}) inversion problems. However, these suffer from high computational costs, which scales with the complexity of the source model. This limits the complexity of the problems that are viable to compute and the usability of the method in the field, where results need to be obtained within minutes. In this work, a \acrshort{ns} method is applied as a benchmark algorithm to evaluate the relative accuracy and speed of the \acrshort{nf} inversion presented.

\subsection{Normalising Flows}\label{normalisingflows}
\acrshort{nf}s are a type of generative neural network that are ideally suited for the modelling and sampling of complex data distributions. The basis of the method described in \cite{Papamakarios(2021)} is to express the data vector ($\boldsymbol{\theta}$) as a transformation ($T$) of another vector ($\boldsymbol{u}$), which is sampled from an arbitrary \textit{base distribution} ($p(\boldsymbol{u})$). The \acrshort{pdf} of $\boldsymbol{\theta}$ can then be defined via the change of variables formula
\begin{equation}
\label{eqn:changeofvar}
p(\boldsymbol{\theta})=p(\boldsymbol{u})\det|J_T(\boldsymbol{u})|^{-1},
\end{equation}
where $J_T$ is the Jacobian matrix containing the partial derivatives of $T$ with respect to $\boldsymbol{u}$. To be able to model rich and complex \acrshort{pdf}s, $T$ can be a composite transformation made up of a number of invertible and differentiable functions. These stacked transformations are often implemented as a neural network with trainable parameters ($\boldsymbol{\phi}$). The network parameters are iteratively tuned during training with the aim to minimise the loss function, which is defined as the \acrlong{kl} (\acrshort{kl}) divergence between the network output ($q(\boldsymbol{\theta})$) and the target distribution ($p(\boldsymbol{\theta})$). When the network is fully trained, the output of the last layer of the network after forward propagation, usually referred to as the \textit{latent space} ($q(\boldsymbol{u})$), is equal to the base distribution. Then, samples from the base distribution ($p(\boldsymbol{u}) = q(\boldsymbol{u})$) can be passed through the inverted transformations to obtain a sample from the real data space and the posterior function can be evaluated. Since $p(\boldsymbol{u})$ can be freely chosen, it is beneficial to opt for a distribution which can be cheaply sampled from, and it is often set to be a multi-dimensional unit Gaussian.

The algorithm described above is a generative model which can be used to interpolate a discrete data set and generate samples from the resulting continuous distribution. However, to apply the network to parameter inference, a {\it conditional} input needs to be introduced alongside the training data. Assume that the aim is to infer the posterior \acrshort{pdf} of some parameters ($\boldsymbol{\theta}$) conditioned on some observation ($\boldsymbol{d}$). Then, the \acrshort{nf} can be trained to approximate the conditional posterior \acrshort{pdf} ($p(\boldsymbol{\theta}|\boldsymbol{d},A)$), such that 
\begin{equation}
\label{eq:conditionalflow}
p(\boldsymbol{\theta}|\boldsymbol{d},A)\approx q(\boldsymbol{\theta}|\boldsymbol{d},A) = q(\boldsymbol{\theta};\boldsymbol{\phi}|\boldsymbol{d}) = p(\boldsymbol{u|\boldsymbol{d}})|\det J_{T}(\boldsymbol{u};\boldsymbol{\phi})|^{-1}
\end{equation}
where $q(\boldsymbol{\theta};\boldsymbol{\phi}|\boldsymbol{d})$ is the output of the trained neural network, $T$ denotes the transformations parameterised by the network parameters $\boldsymbol{\phi}$ and $A$ is an assumed model for the data.

To apply this method to gravity inversion, a data set of source parameters ($\boldsymbol{\theta}$) needs to be generated from the chosen prior and the corresponding gravity measurements ($\boldsymbol{d}$) can be simulated via the gravity forward model ($A$), which forms the conditional part of the data set once simulated noise is added. Once the network is trained on this data set, samples can be generated from the posterior distribution of source parameters conditioned on a specific gravimetry survey of interest, only returning parameters that are consistent with the measurements. The \acrshort{nf} method offers a significant improvement in terms of computation time as compared to \acrshort{ns}, as once trained, inversion results can be obtained within seconds and the trained network can be reused for multiple surveys.

\subsection{Forward model}\label{forwardmodel}
In gravity modelling, the forward model is the function that describes the one-to-one mapping from the underground density distribution to the theoretical surface gravity measurements. In traditional inversion methods iterative forward calculations take place and the calculated surface gravity is compared to the true values for optimization. In a \acrshort{ns} algorithm iterative calculations are also required, as the forward model forms part of the likelihood function defined in Section \ref{implementation}. When applying a generative neural network for inversion, we are also required to define a forward model, but these are calculated in advance to form a part of the training data set. In this work an analytical forward model is applied, which is specific to rectangular prism anomalies (\cite{Li(1998)}).

\begin{figure}[]
\centering
\includegraphics[width=10cm]{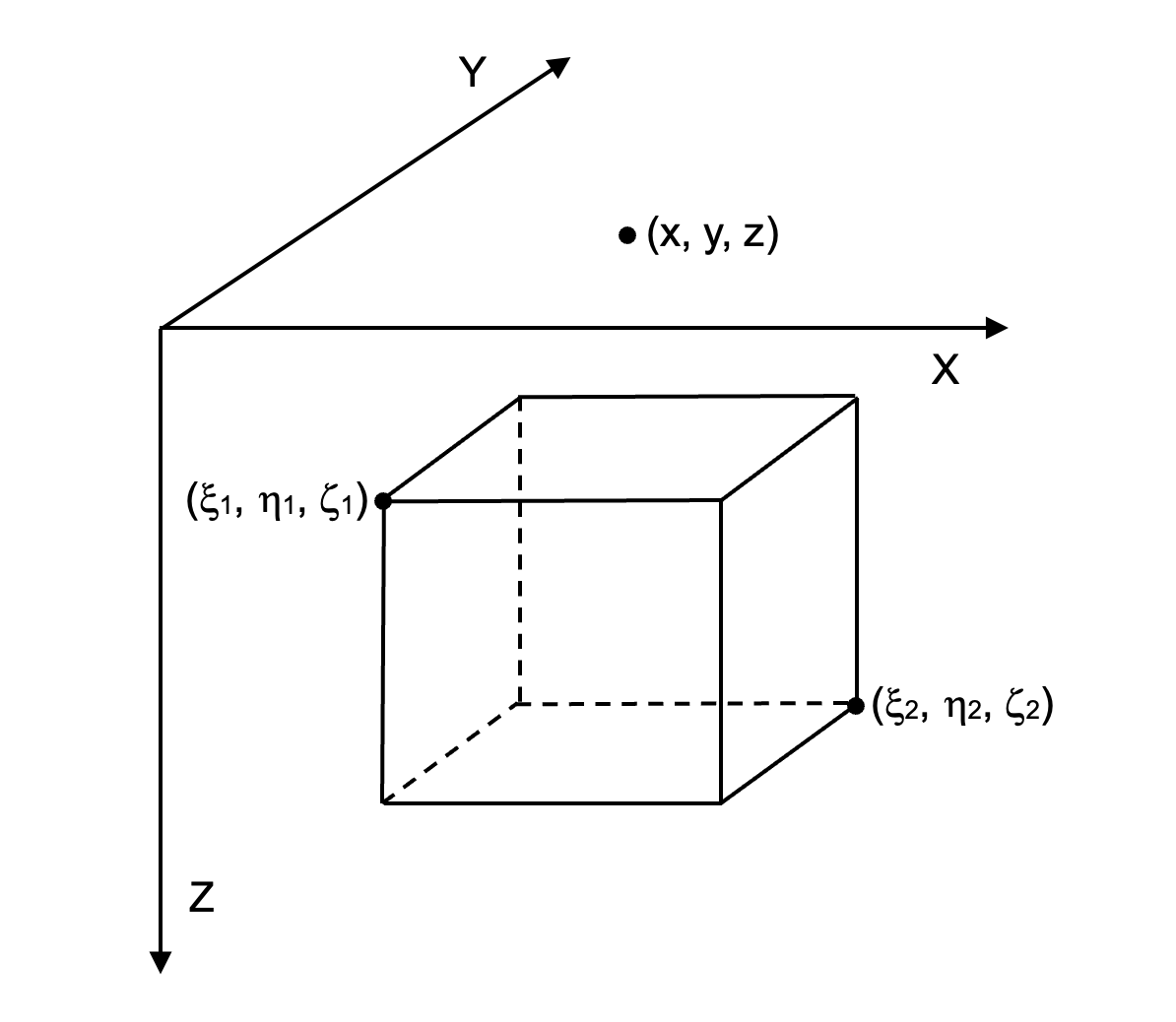}
\caption{Diagram of the rectangular prism density anomaly for which the forward model of the gravity measured at the point $(x, y, z)$ is defined in Equation \ref{eq:forwardmodel1}. (\cite{Li(1998)})}
\label{fig:prismdiagram}
\end{figure}

The prism, visualised in Figure \ref{fig:prismdiagram}, is defined to have its faces aligned with the $X$, $Y$ and $Z$ axes, with limits $[\xi_1, \xi_2]$, $[\eta_1, \eta_2]$ and $[\zeta_1, \zeta_2]$ correspondingly. The vertical component of the gravitational acceleration ($g$) at the location $R = (x, y, z)$ due to this body is
\begin{equation}
\label{eq:forwardmodel1}
g(R; \boldsymbol{\theta}) =  G\rho\sum_{i=1}^2\sum_{j=1}^2\sum_{k=1}^2\mu_{ijk}\left[x_i\ln(y_j+r_{ijk})+y_j\ln(x_i+r_{ijk})-z_k\arctan\frac{x_iy_j}{z_kr_{ijk}}\right]
\end{equation}
where
\begin{equation}
\label{eq:forwardmodel2}
x_i=x-\xi_i, y_j=y-\eta_j, z_k=z-\zeta_k, \\
r_{ijk}=\sqrt{x_i^2+y_j^2+z_k^2}, \\
\mu_{ijk}=(-1)^i(-1)^j(-1)^k,
\end{equation}
$G$ is the gravitational constant and $\rho$ is the density of the object. The source parameters ($\boldsymbol{\theta}$) that were chosen to describe each instance of the rectangular prism anomalies are listed in Table \ref{tab:priors}, and it is discussed in Section \ref{testcases} how these are translated into the prism limits ($\xi, \eta, \zeta$).

\section{Test Cases}\label{testcases}
The performance of the inversion algorithm was tested on a simulated data set of source models consisting of rectangular prism voids located underneath the surface. This data set was chosen as a simple test case that can be described by a handful of parameters and for which the results are straight-forward to visualise and interpret. While the modelling of natural geological features usually requires more complex models, a simple data set as such can be applied to the modelling of some human-made architecture, such as tunnels or bunkers. The network was tested on two different data representations to showcase the performance on data with varying dimensionality.

In the low-dimensional (parameterised) case, the location and size of the rectangular prisms were described by 7 parameters, listed in Table \ref{tab:priors} alongside the chosen prior \acrshort{pdf}s. To generate the training set, 1 million samples were drawn from the prior and the forward model was calculated for each sample following the method described in Section \ref{forwardmodel}. For this, the source parameters can be translated into the prism limits ($\xi, \eta, \zeta$) in Equation \ref{eq:forwardmodel2} via eg., $\xi_1 = c_x - l_x/2$ and $\xi_2 = c_x + l_x/2$.  Since for the specified forward model the faces of the prism need to be aligned with the $X, Y$ and $Z$ axes, the rotation parameterised by $\alpha$ is applied to the survey coordinates in the opposite direction, resulting in the same effect in the gravity values. The density contrast between the background and the object itself ($\rho$) was set to 1500 kg/m$^3$, approximated from the assumption that it is a hollow prism encased in topsoil. The comparison between the \acrshort{nf} and \acrshort{ns} algorithms is only done for this test case.

In the higher-dimensional (voxelised) case, the source parameters were first drawn from the same priors as before and were translated into a 512-dimensional array of densities of cubic voxels on an $8\times8\times8$ grid covering a region of 120 m in each direction. This was done by calculating the fraction of each voxel that overlaps with the rectangular prism, and assigning a density value according to the average of the background and the object densities weighted by this fraction. In addition, Gaussian noise with a standard deviation of 50 kg/m$^3$ was added to the density values, to simulate fluctuations in topsoil density. Each voxel is then considered as an instance of a rectangular prism, for which the analytical forward model is calculated, and finally the results are summed for all voxels. Similar to the parameterised case, the prism limits for Equation \ref{eq:forwardmodel2} can be computed from the coordinates of the centre and the size of each voxel. The rotation angle $\alpha$ is accounted for during the translation of the prism parameters into density values, therefore the survey itself is not rotated.
While in this case swapping to this data representation seems impractical as it reduces the resolution of the model while increasing the data volume and introducing broader priors, it serves as a stepping-stone towards the modelling of more complex non-rectangular bodies.

For the survey simulations, the stations were set to be arranged on an 8$\times$8 uniform grid with a separation of 10 m in each direction, and the elevation kept at a constant. The survey was chosen to cover a smaller area that the voxel grid to avoid edge effects, which means that the prism is allowed to be located outside of the survey region. Since in real gravimetry surveys taken in the field it is difficult to take measurements on an exact grid, the voxelised data set was also combined with a noisy survey configuration. Gaussian noise with a standard deviation of 2 $m$ was added to the $x$ and $y$ coordinates of the stations. In the case where the survey locations do not change between examples in the data set, only the gravity values were provided as the conditional input to the \acrshort{nf}. However, in the case of varying, noisy survey locations, the $x$ and $y$ coordinates for each measurement were also included. Overall three different test cases were considered; the parameterised data representation with survey measurements arranged on a uniform grid, and the voxelised data representation with both a uniform survey grid and one with added noise to the grid locations. For all cases Gaussian noise with a standard deviation of 10 $\mu$Gal was also added to the survey measurements to simulate instrumental noise, which was chosen as an arbitrary small value compared to the magnitudes of the expected relative gravity signals.

\section{Implementation}\label{implementation}
The benchmark \acrshort{ns} method was implemented using built-in functionalities in the {\tt bilby} software package (\cite{bilby}), using the {\tt dynesty} algorithm (\cite{dynesty}). This method requires the definition of prior \acrshort{pdf}s, which are defined in Table \ref{tab:priors} for the 7 source parameters. An analytical likelihood function also needs to be specified, which is used to model the noise on the measurements. For the purposes of this work, Gaussian noise with a standard deviation of $\sigma=$ 10 $\mu$Gal was added to the simulated gravimetry data, therefore a Gaussian likelihood function needs to be implemented. For a set of proposed source parameters ($\theta$) this is described as
\begin{equation}
\label{eqn:likelihood}
    \mathcal{L}(d|R,\boldsymbol{\theta}) =  \prod_{i} \frac{1}{\sqrt{2\pi \sigma^2}}\exp{\left(\frac{-(d_i-g(R_i;\boldsymbol{\theta}))^2}{2\sigma^2}\right)}, \\
    d_i = g(R_i;\boldsymbol{\theta_0}) +n_i
\end{equation}
where it is assumed that $d_i$ are independent data points measured at locations $R_i$. The data is assumed to consist of a forward model $g(R_i;{\boldsymbol{\theta}})$, which is described in Section \ref{forwardmodel} and computed from the true source parameters ($\boldsymbol{\theta_0}$), and noise ($n_i$) sampled from a Gaussian distribution with standard deviation $\sigma$. This is then compared to the forward model computed from the sampled source parameters ($\boldsymbol{\theta}$) at each iteration of the algorithm and the total likelihood is found from the product of each component at locations $R_i$.

\begin{table}[]
\centering
  \caption{The 7 parameters that define the location and size of the prism. The prism can only rotate around the z axis centred on $(c_x,c_y)$, and the rotation angle is defined as $\alpha$, which is constrained to $<\pi/2$. If the prism were allowed to rotate by more than $\pi/2$, the same configuration could be reproduced by taking a prism and rotating it by the angle $\pi/2+\alpha_0$ and by taking a prism with its $l_x$ and $l_y$ coordinates swapped and rotating it by $\alpha_0$, hence introducing a degeneracy in the posterior. In addition, the limits on the location and size in the $z$ dimension enforce that the prism remains below the surface.}
 \label{tab:priors}
 \begin{tabular}{|c || c| c|} 
 \hline
 Parameter & Description & Prior \\ [0.5ex] 
 \hline \hline
 $c_x$ & Centre of the prism in the $x$ dimension. & Uniform(-60,60) \\ 
 \hline
 $c_y$ & Centre of the prism in the $y$ dimension. & Uniform(-60,60)\\
 \hline
 $c_z$ & Centre of the prism in the $z$ dimension. & Uniform(-60,20)\\
 \hline
 $l_x$ & Length of the prism in the $x$ dimension. & Uniform(0,120) \\
 \hline
 $l_y$ & Length of the prism in the $y$ dimension. & Uniform(0,120)\\
 \hline
 $l_z$ & Length of the prism in the $z$ dimension. & Uniform(0,80) \\
 \hline
 $\alpha$ & The rotation angle around the z axis. & Uniform(0,${\pi}/{2}$) \\ [1ex]
 \hline
 \end{tabular}
\end{table}

The \acrshort{nf} method was implemented using {\tt glasflow} (\cite{glasflow}), a python package based on {\tt pytorch} (\cite{pytorch}) and {\tt nflows} (\cite{nflows}). Before training, the data set consisting of 1 million data points was split into 90\% training and 10\% validation data and was normalised using the {\tt scikit-learn MinMaxScaler} tool, which scales the data separately for each dimension so that the values lie within the $[0, 1]$ range. This was done for both the data set of source models and the conditional data consisting of survey measurements. The chosen architecture is built up of RealNVP (\cite{Dinh(2017)}) transformations and is described by the parameters listed in Table \ref{tab:networkparameters}. The final network parameters were selected based on rigorous optimisation using {\tt Weights \& Biases} (\cite{wandb}) hyper-parameter optimisation tools. The network architecture was optimised for the parameterised data set, which was tested most rigorously, and was kept unaltered for the remaining test cases to promote the flexibility and user-friendliness of the method.

The network was trained using the {\tt pytorch} built-in {\tt Adam} optimiser with an initial learning rate of 0.001 and a {\tt ReduceLROnPlateau} learning rate scheduler, which reduces the learning rate by a factor of 0.05 whenever the validation loss stops decreasing for 100 epochs. The training was completed when the learning rate reached $10^{-6}$ and the validation loss reached a plateau. Training can be done without the implementation of learning rate scheduler, however this can make the process more efficient and allows the user to utilise various learning rate values during a single training run. The network architecture was constructed so that it can remain unchanged for the three different test cases, however the required training time varied from $\sim$4.5 hours for the parameterised case to $\sim$13 hours for the voxelised case with a noisy survey grid using a NVIDIA Geforce RTX 3090 GPU. This increase in training time is largely due to the increase in data volume. Once the network is trained, 1000 samples can be obtained within ~7 seconds as compared to the $\sim$20 minute computation time associated with a single \acrshort{ns} run, and the trained network can be reused for multiple inversion cases. The whole pipeline with examples is available on GitHub (\cite{flowinv}).

\begin{table}[]
\centering
  \caption{The parameters that define the RealNVP (\cite{Dinh(2017)}) neural network architecture that was implemented in this work. Only the parameters that differ from the default values implemented in {\tt glasflow} (\cite{glasflow}) are listed here.}
 \label{tab:networkparameters}
 \begin{tabular}{|l || c| } 
\hline
  Parameter & Value \\ [0.5ex] 
 \hline \hline
 Transforms & 12 \\
 \hline
 Blocks per transforms & 2 \\ 
 \hline
 Neurons &  64 \\
 \hline
 Batch size & 5000 \\
 \hline
 Initial learning rate & 0.001 \\
 \hline
 \end{tabular}
\end{table}

\section{Results}\label{results}
In this section the results for the gravity inversion using the three different data set discussed in Section \ref{testcases} are discussed. The devised testing methodologies are outlined and are used to scrutinise the performance and accuracy of the presented \acrshort{nf} inversion method, and comparisons are made with the \acrshort{ns} wherever relevant.

\subsection{Survey consistency test}\label{surveyconsistency}
\begin{figure}[]
\centering
\includegraphics[width=10cm]{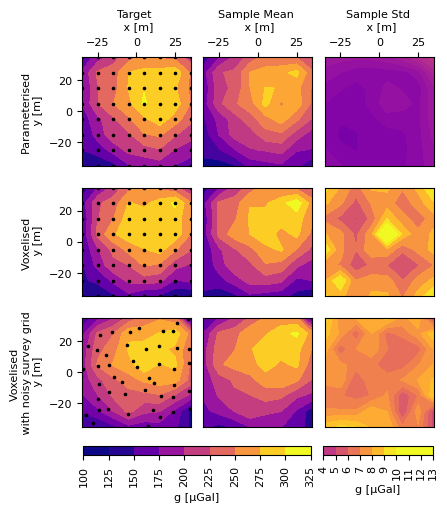}
\caption{Plot showing the comparison between the true gravimetry survey and the simulated surveys from the \acrshort{nf} output. Samples were drawn from the posterior distribution of source parameters produced by the 3 different test cases, and each were forward modelled using the method outlined in Section \ref{forwardmodel}. Column 1 is the input gravimetry survey, column 2 is the pixel-wise mean computed from 2000 samples and column 3 is the standard deviation.}
\label{fig:surveyconsistency}
\end{figure}

When using the trained network for inversion, samples are drawn from the posterior probability distribution over the source parameters, conditioned on the specific gravimetry survey of interest. The samples represent individual instances of the source parameters that are all consistent with the survey data within the limits of instrumental noise. To test this consistency, a number of samples can be drawn from the posterior distribution of the \acrshort{nf} and forward modelled to compare to the original survey data. Summary statistics based on 2000 samples are shown in Figure \ref{fig:surveyconsistency} for the three separate test cases. The pixel-wise mean of the forward-modelled samples aligns well with the input survey and the standard deviation is $\sim$1\% of the amplitude of the relative gravity signal for the parameterised case, and $\sim$5\% for the voxelised cases. This difference can be attributed to the relative difficulties of the different test cases due to higher dimensionality and wider prior spaces, as well as to the additional noise contributions from the background density fluctuations, which are not present in the parameterised data. These results demonstrate that the samples represent valid solutions to the inversion problem.

\subsection{Statistical consistency test}\label{statisticalconsistency}
Another important test to conduct is to prove that the resulting posterior from the sampling method is statistically consistent and free of bias. The probability-probability (p-p) plot is a standard method developed to validate Bayesian algorithms (\cite{Cook(2006)}, \cite{Talts(2020)}). For this method, probability is considered as defined by frequentist statistics, where the interest is the frequency ($F$) at which the true parameter value lies within the $F\%$ credible interval of the posterior probability distribution. In frequentist probability tests having a sufficiently high number of test cases is essential, however a trade-off with computational cost needs to be considered. In this work, 100 test cases were randomly chosen from the prior to produce each p-p plot.

\begin{figure}[]
\centering
\includegraphics[width=16cm]{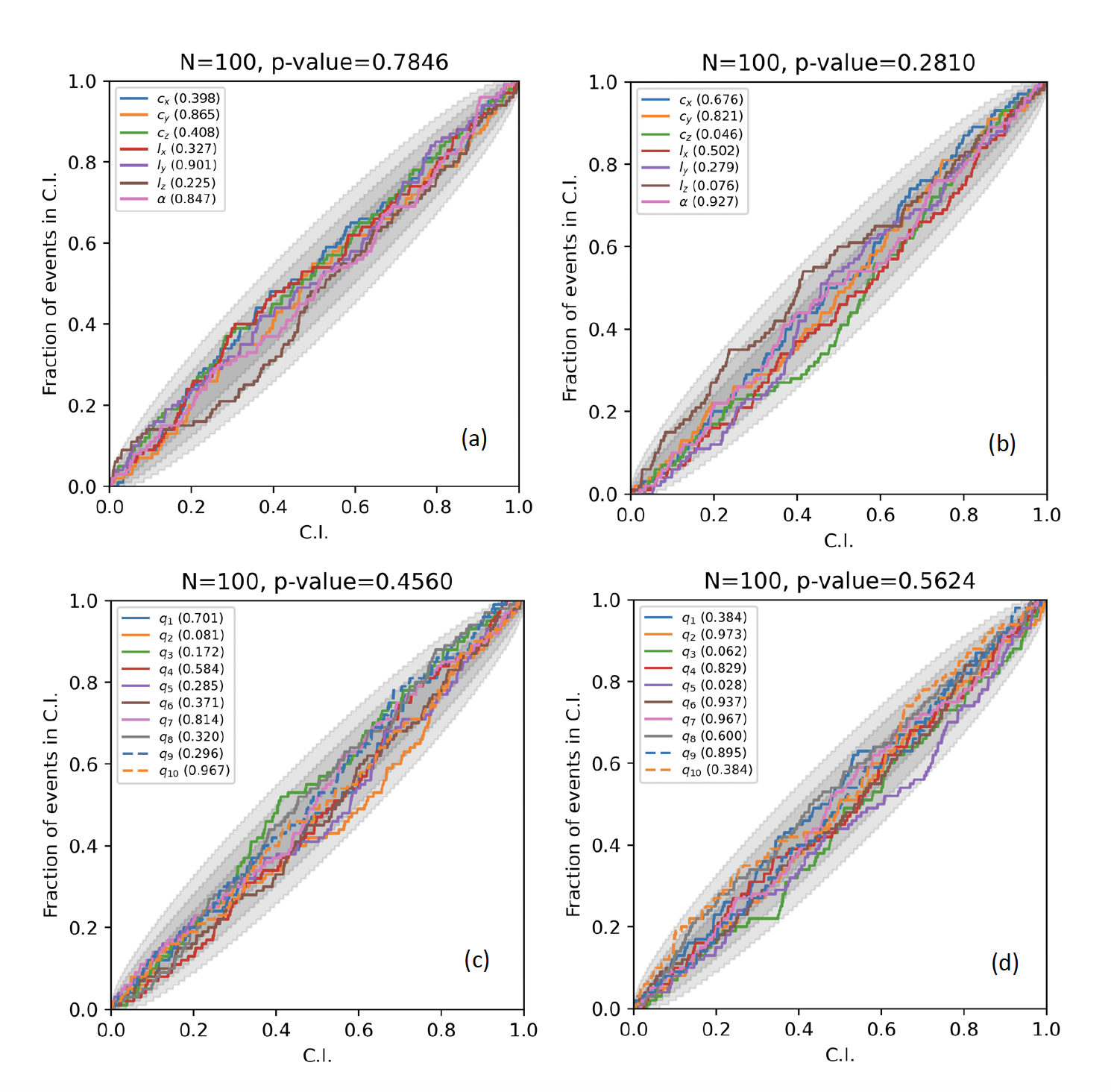}
\caption{The various p-p plots obtained from the two different inversion methods and three data sets; (a) Nested Sampling inversion of the parameterised data set, (b) Normalising Flows inversion of the parameterised data set, (c) Normalising Flows inversion of the voxelised data set and (d) Normalising Flows inversion of the voxelised data set with added noise to the survey locations. For the voxelised test case only a randomly picked 10 parameters \{$q_1 ... q_{10}$\} of the 512 dimensions were plotted for better visibility. The p-values are displayed in the legend for each curve separately, and the combined value is displayed above each graph. The method of obtaining these plots is described in detail in Section \ref{statisticalconsistency}.}
\label{fig:pp_plot}
\end{figure}

To construct the p-p plot shown in Figure \ref{fig:pp_plot}, the following process was applied; first 100 sets of source parameters were generated from the prior and were forward modelled to compute the simulated gravity surveys with Gaussian noise added to the measurements. Then, a number of samples were drawn conditioned on each of these surveys separately from the posterior distribution of source parameters output by the inference method. The marginalised \acrshort{cdf} at the location of the true parameter value in each dimension was then calculated for each test case separately. A histogram of the values of the obtained \acrshort{cdf}s at the true parameter locations was computed, which approximates to a uniform distribution when the method is consistent. Finally, the \acrshort{cdf} of this distribution was plotted, resulting in a series of lines following the diagonal in the ideal case. However, due to the relatively low number of test cases considered when constructing this plot, slight deviations from the diagonal are expected. The method is considered consistent as long as the curves lie within the 68\%, 95\% and 99.7\% confidence intervals, which are marked by the grey regions on the plot. When significant deviations from the diagonal are observed, the shape of the curve can hint at specific bias in the posterior. For example, if an S shape is observed, the posterior is over- or under-constraining the parameter space, while in case of an bow shape, the posterior is shifted from its ideal location.

The $p$-value was also computed from the maximum distance between the diagonal line and the curves corresponding to the various parameter dimensions using the Kolmogorov-Smirnov test. It is the probability of obtaining a value of a distance statistic, or greater, under the null hypothesis that the underlying cumulative distribution is diagonal. Extremely small $p$-values indicate that the null hypothesis is unlikely. The combined value displayed on the top is obtained using Fisher's method to compute the $p$-value where the null hypothesis is that each parameter obeys its own null hypothesis. The $p$-values for each curve as well as the combined value are all expected to appear to be random samples from a uniform distribution between 0 and 1. Figure \ref{fig:pp_plot} shows evidence of the statistical consistency of the \acrshort{ns} and the various \acrshort{nf} inversion test cases.

\subsection{Parameterised posterior comparison}\label{parameteraccuracy}
\begin{figure}[]
\centering
\includegraphics[width=17cm]{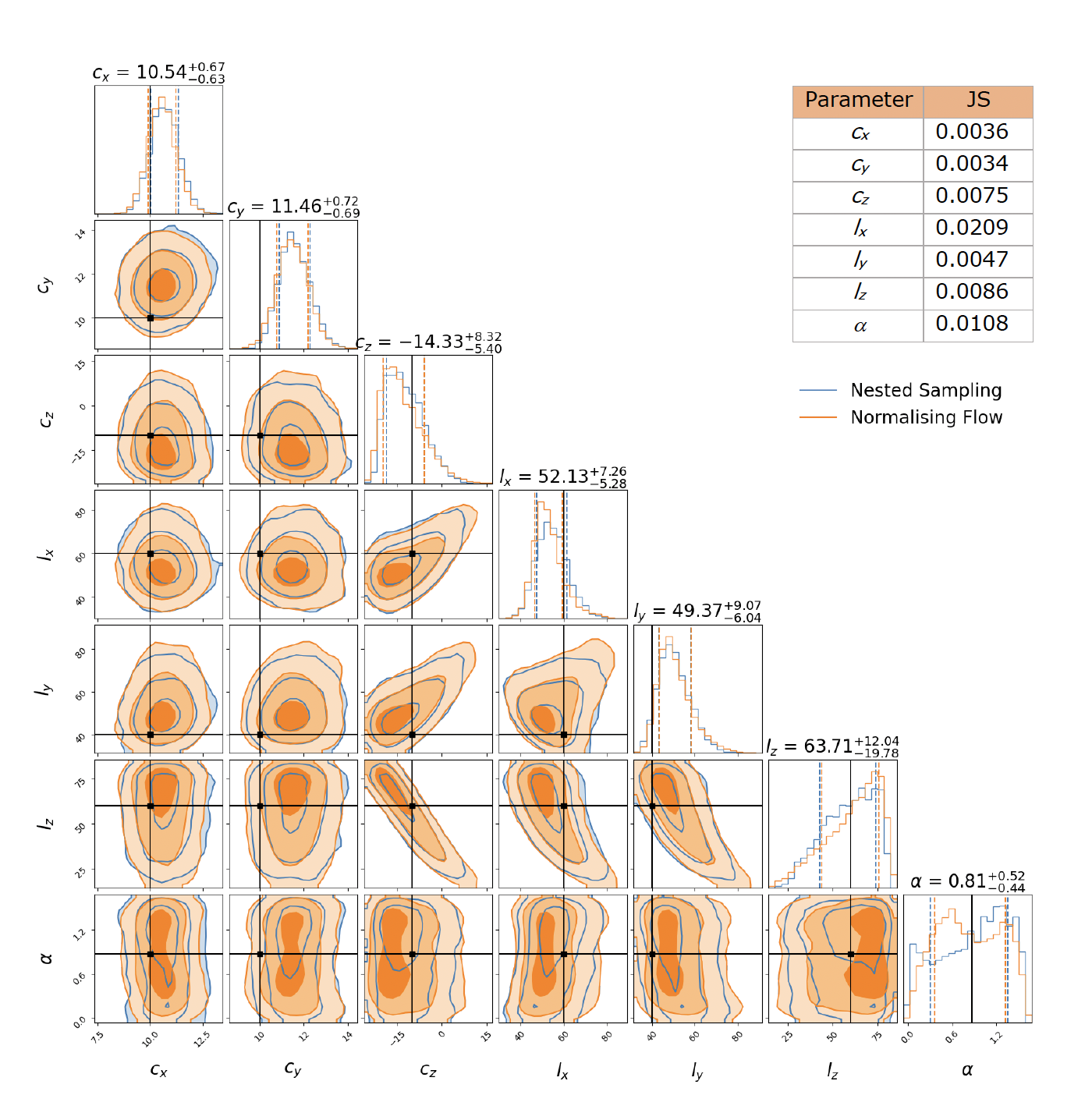}
\caption{The corner plot comparing the posterior probability distributions obtained from the Nested Sampling (blue) and the Normalising Flows (orange) for the specific test case. The 1-dimensional marginalised probability distributions are shown on the diagonal for each parameter separately, with the mean of the posterior distribution obtained from the Normalising Flows displayed above. The dashed lines and the error bounds correspond to the 16$\%$ and 84$\%$ quantiles of the distribution. The black lines and markers correspond to the true parameter values. The method of obtaining this plot is explained in detail in Section \ref{parameteraccuracy}. On the right hand side a table of the Jensen-Shannon (JS) divergence between the 1-dimensional marginalised posterior distributions produced by the Nested Sampling and Normalising Flows methods are shown for this test case.
}
\label{fig:cornerplot}
\end{figure}

We are also interested in visualising the posterior \acrshort{pdf} of source parameters, and testing the accuracy of the inference from the \acrshort{nf} compared to the \acrshort{ns} method. In the case of a handful of parameters, the results of the inference can be visualised on a so-called corner plot. This array of graphs depicts the marginalised 1-dimensional probability distributions of individual parameters on the diagonal panels, and the 2-dimensional combined distributions of pairs of parameters in the panels below the diagonal. The locations of the true parameter values were marked for reference, which are expected to be consistent with the posterior contours. This plot can also be used to compare probability distributions obtained from different inference methods applied to the same input data under the same model and prior assumptions. If both methods perform well, it is expected that these would constrain the parameter space at the same level of accuracy, and ideally would be identical within the uncertainties of finite sampling. The results for the inversion of the gravimetry survey in the first row and first column of Figure \ref{fig:surveyconsistency} are shown in Figure \ref{fig:cornerplot}.

Some interesting correlations can be observed in the arising distributions. First of all, there is a negative correlation between the extent in the $z$ dimension ($l_z$) and depth of the centre of the prism ($c_z$). This can be contributed to the fact that the same gravity anomaly can be linked to a family of solutions ranging from a small and shallow object, to a larger object deeper underneath the surface. From vertical gravitational acceleration data only, these are not possible to distinguish. However, other less obvious correlations appear as well. Both the $c_z$ and $l_z$ parameters are correlated with the $l_x$ and $l_y$ parameters. This can be explained by the angle between the gravity vector pointing to each point along the edge of the prism in the $x-y$ plane and the vertical component of this vector remaining constant. Since the density of the prism is constrained, the contribution of each point within the prism to the total gravitational acceleration vector at each survey location is also constrained. Since the vertical component of gravity is directly measured, for the survey measurements to remain unchanged for prisms located at different depths, the angle between the gravity vector and its vertical component has to remain the same. Therefore, when considering survey locations just outside the boundaries of the prism, the edge of the prism has to lie closer to these points when closer to the surface, and further away when deeper beneath the surface, effectively shrinking the prism in the $x$ and $y$ dimensions. This effect can be visualised using the analogy of shining a torch onto a screen, placing a box in-between them and observing the shadow cast onto the screen. When the box is closer to the screen (the prism is closer to the surface), its image will be closely aligned with its size. However when the box is moved towards the torch (the prism is deeper beneath the surface), its shadow becomes larger.

\begin{figure}[]
\centering
\includegraphics[width=8cm]{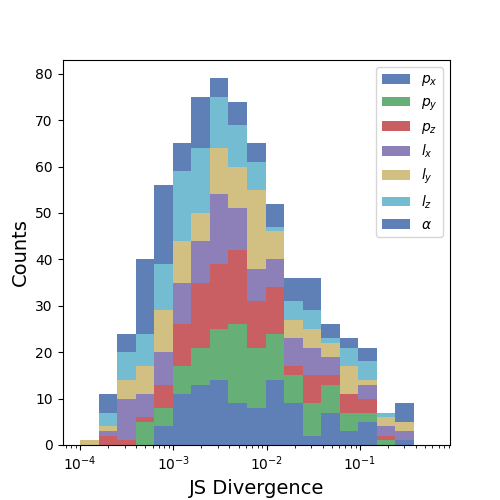}
\caption{
A histogram of the Jensen-Shannon (JS) divergence between the 1-dimensional marginalised posterior distributions of source parameters produced by the Nested Sampling and Normalising Flows methods for 100 test cases. This provides a measure for the distance between the posterior distributions from the Normalising Flow and the assumed ideal distribution, which is obtained from the Nested Sampling method. The range of possible values for the JS divergence is $[0, \ln(2)]$, with larger values indicating larger differences between the distributions. The number of counts of JS divergence values within each bin for the different parameter dimensions are marked by different colours and are stacked on top of each other to show and overall distribution. Note the logarithmic scale on the $x$-axis for better visibility.
}
\label{fig:jshist}
\end{figure}

While the corner plot in Figure \ref{fig:cornerplot} shows that the results from the two methods agree and have comparable accuracy for a single test case, it is also necessary to understand the general performance of the \acrshort{nf}. In Figure \ref{fig:jshist} a histogram of the \acrlong{js} (\acrshort{js}) divergence between the 1-dimensional marginalised posterior \acrshort{pdf}s obtained from the two different methods is shown. This provides a measure of distance between the distributions, with a minimum value of 0 and a maximum of $\ln{2}$. It is shown here that 98\% of the distributions have a \acrshort{js} divergence $<0.1$ with a median of 0.0039. These results indicate that the \acrshort{nf} can consistently produce accurate posteriors assuming that the \acrshort{ns} approach produces the true posterior. One of the potential reasons explaining the rare cases of higher JS divergence values is the emergence of topologies in the true posterior \acrshort{pdf} which are difficult to model using \acrshort{nf}s, such as multi-modality (\cite{Dinh(2020)}). In a few specific cases when the $\alpha$ parameter lies close to the edge of the prior space, the posterior of $\alpha$, $l_y$ and $l_x$ become multi-modal. This is due to a specific degeneracy in the formulation of the prior, where a prism rotated by $\alpha=\pi/2$ with specific $l_x$ and $l_y$ dimensions would look the same as a prism with $\alpha=0$ with its dimensions swapped around.

\subsection{Density distribution comparison}\label{densitydistribution}
\begin{figure}[]
\centering
\includegraphics[width=16cm]{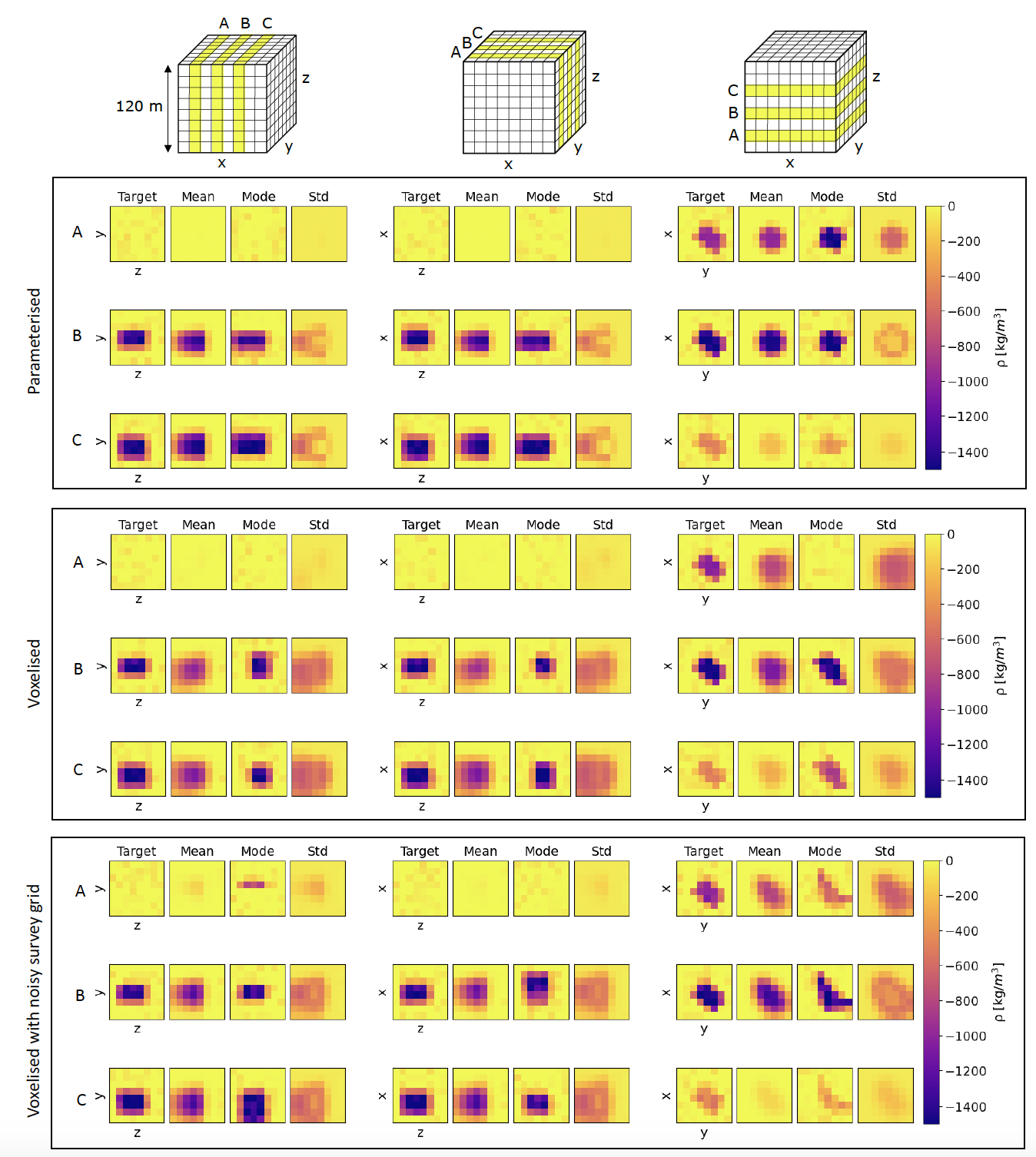}
\caption{Plot comparing the output of the Normalising Flows inversion to the target density distribution. The three subplots correspond to the three different test cases; the parameterised data set, the voxelised data set, and the voxelised data set with added noise to the survey locations, from top to bottom respectively. Each column corresponds to the $x$, $y$ and $z$ dimensions in order along which the volume is sliced, and the subcolumns in order are the target distribution, and the mean, mode and standard deviation of the samples from the posterior distribution of source parameters output by the Normalising Flow. The rows correspond to the different slices, as marked on the diagrams on top.}
\label{fig:densitycompare}
\end{figure}

In case of the voxelised source model representation, each sample from the \acrshort{nf} is a 512-dimensional array of parameters representing the voxel densities. Due to the high dimensionality, it is inconvenient to directly visualise the posterior on a corner plot, and it is more informative to compute summary statistics about the distribution. The mean, mode and standard deviation of 10000 samples from the posterior distribution of the source parameters are shown in Figure \ref{fig:densitycompare} alongside the target source model, which is a single piece of data the method is tested on. While in these test cases the source model from which the gravimetry survey was simulated is known, it is not expected that the inversion would return the true model when only the survey data is provided due to the inherent degeneracies and noise added during the simulation. It can however be expected that the resulting distribution of source parameters includes the true model. To allow comparison between the results from the parameterised and the voxelised inversion, the results from the parameterised case were also translated into density values and the same summary statistics were plotted in the topmost subplot of Figure \ref{fig:densitycompare}.

The mean and standard deviation of the distribution of the source parameter vectors were computed element-wise. The mean provides a visual representation of all the locations where the prism is potentially located, and the standard deviation is indicative of the uncertainty in the density predicted for the specific voxel, which is highest around the predicted edges of the prism. The mode is found by choosing the sample from the \acrshort{nf} that has the highest probability associated with it, hence it marks the single solution approximating the peak of the multi-dimensional posterior. This is not expected to closely agree with the true source model, it is only the most probable solution based on the data. This phenomenon is illustrated in the bottom subplot of Figure \ref{fig:densitycompare}, where the mode of the distribution disagrees with the true model in terms of its extent and location in the $z$ dimension as well as its size in the $y$ dimension, while in the top subplot the mode agrees with the truth more closely. In addition, the correlations observed in Section \ref{parameteraccuracy} appear in Figure \ref{fig:densitycompare} as a cone-like feature in the mean of the slices along the $x$ and $y$ dimensions, with the cone getting narrower further from the surface. Therefore, the deeper the prism is, the smaller it has to be in the $x$ and $y$ dimensions to remain consistent with the gravimetry survey data. This is most apparent in the bottom panel presenting results from an inversion with a noisy survey grid. In general, the uncertainty in the results for this case appears lower than for the inversion of the survey taken on a uniform grid, which suggests that a more randomised survey configuration is more ideal for the \acrshort{nf} inversion method.

\section{Discussion}\label{discussion}
There are various aspects that determine how well the parameter space can be constrained for a specific inverse problem. Firstly, the distribution is broadened by the inherent non-uniqueness which is ubiquitous in inverse problems in gravimetry. The method presented here has the advantage over conventional single-solution inversion that it does not ignore this non-uniqueness and incorporates all possible solutions in the results. More importantly, the resolution of the gravimetry survey and the instrumental noise have a large effect on the accuracy with which the source parameters can be inferred. These effects cannot be removed and can only be reduced by intelligent survey design, good instruments and repeated measurements. However, as the target distribution becomes narrower, the more challenging the inference problem becomes and training time and network complexity increases. The level of noise included in this work is relatively low compared to that of currently available gravimeters on the market and only a simple Gaussian noise profile was implemented here. When designing a model for surveys taken with a specific gravimeter, more informed instrumental noise profiles can be applied.

One benefit of using machine learning for inversion is the speed at which results can be obtained once the network is trained. However, this feature can only be taken advantage of if the trained network is reusable, which means that the chosen priors need to be general enough to be applicable to a range of gravimetry surveys, but restrictive enough not to return irrelevant results. A simple uniform prior space was implemented here, but for real-life applications it is recommended to implement more restrictive priors with as much information content as possible to make the inference faster and more accurate. In addition, increasing model complexity can make the data set more applicable to real-life situations. Any feature that can be parameterised and its gravitational effect can be modelled, such as the geological background or previous knowledge gained from other geophysical surveys, can be incorporated in the source models. However, with increasing model complexity the volume of the training data required to span the prior space is likely to increase too. Depending on the speed of the forward model, data generation can prove to be a bottleneck and a trade-off between the advantages of more complex models and the computation time needs to be considered. The performance of the \acrshort{nf} was only tested for up to a few hundred data dimensions. Some inversion targets might require even higher complexity to model, therefore exploring the performance at higher dimensions and combining the inference with data reduction methods are desirable avenues of research.

Another restriction in the current configuration of the method is that the number of points in the gravimetry survey needs to be decided before the measurements are taken and kept at a constant across different surveys if the trained network is to be reused. For more flexibility, it would be beneficial to develop the method to be able to take in a variable number of survey points at unrestricted locations. This would also better allow the inversion to be used in the field, for which the \acrshort{nf} method is ideally suited due to fast computation times. It could be used to immediately obtain inversion results, to guide the user towards the best survey locations to be measured and to indicate when the desired uncertainty threshold was found and the survey can be concluded.

\section{Conclusions}\label{conclusions}
Bayesian inference methods have been successfully applied to gravity inversion, however standard sampling methods suffer from long computation times, which increases with the dimensionality of the data set. Here, a gravity inversion method was implemented using Normalising Flows, which can improve on standard sampling methods in terms of computation time and flexibility. First, the method was tested with a data set consisting of rectangular prism bodies described by 7 source parameters. The survey measurements were simulated using the forward model described in \cite{Li(1998)} and Gaussian noise was added to the measurements to simulate instrumental noise. The performance of Normalising Flows for this low-dimensional example was compared to that of Nested Sampling, and it was shown that the accuracy is comparable, while the Normalising Flow can produce results within $\sim$ 7 seconds instead of $\sim$ 20 minutes. To test the network with a higher-dimensional data set, the rectangular prisms were translated into densities defined on a uniform 8$\times8\times$8 cubic grid. It was shown that the posterior distributions of the source parameters remain tractable and accurate results are produced. The method was tested with two different survey configurations, first with a simple uniform 8$\times$8 grid followed by one with added Gaussian noise to the grid locations to more closely simulate realistic survey configurations. As future work, we aim to test the network with more complicated noise profiles and a more flexible survey configurations.

\section{Acknowledgments}

The authors would like to extend their gratitude to Dr. Michael Williams and Christian Chapman-Bird for their invaluable expert advice and insightful ideas, which greatly enriched the contents of this paper. Special thanks go to Mahijs Reinier Koymans and Dr. Craig Whitehill for their revisions and constructive feedback, which played an instrumental role in enhancing the quality of this manuscript. Their contributions have been indispensable in the completion of this research project. The authors acknowledge the usage of the analysis software {\tt nflows} (\cite{nflows}), {\tt glasflow} (\cite{glasflow}), and {\tt bilby} (\cite{bilby}).

This work was primarily funded by UK Research and Innovation (Grant No. ST/V005499/1). CM acknowledges partial support from Science and Technology Research Council (Grant No. ST/L000946/1). For the purpose of open access, the authors have applied a Creative Commons Attribution (CC BY) licence  to any Author Accepted Manuscript version arising.

\bibliographystyle{unsrtnat}
\bibliography{main}  

\begin{thebibliography}{37}
\providecommand{\natexlab}[1]{#1}
\providecommand{\url}[1]{\texttt{#1}}
\expandafter\ifx\csname urlstyle\endcsname\relax
  \providecommand{\doi}[1]{doi: #1}\else
  \providecommand{\doi}{doi: \begingroup \urlstyle{rm}\Url}\fi

\bibitem[Torge(1989)]{Torge(1989)}
Wolfgang Torge.
\newblock \emph{Gravimetry}.
\newblock De Gruyter, 1989.
\newblock ISBN 9783110107029.

\bibitem[Milano et~al.(2021)Milano, Varfinezhad, Bizhani, Moghadasi, Kalateh, and Baghzendani]{Milano(2021)}
Maurizio Milano, Ramin Varfinezhad, Hamid Bizhani, Meysam Moghadasi, Ali~Nejati Kalateh, and Hamidreza Baghzendani.
\newblock Joint interpretation of magnetic and gravity data at the golgohar mine in iran.
\newblock \emph{Journal of Applied Geophysics}, 195:\penalty0 104476, 2021.
\newblock ISSN 0926-9851.
\newblock \doi{https://doi.org/10.1016/j.jappgeo.2021.104476}.
\newblock URL \url{https://www.sciencedirect.com/science/article/pii/S092698512100224X}.

\bibitem[Pearse et~al.(2021)Pearse, Cárdenas~Contreras, Barrera~Lopez, Castillo~Ruiz, Martínez~Gómez, and Tary]{Pearse(2021)}
Jillian Pearse, Andrés Cárdenas~Contreras, Carol~Vanessa Barrera~Lopez, Nataly Castillo~Ruiz, Henry Martínez~Gómez, and Jean~Baptiste Tary.
\newblock Gravity survey and modelling of the nemocón salt mine, colombia.
\newblock \emph{Near Surface Geophysics}, 19\penalty0 (3):\penalty0 365--376, 2021.
\newblock ISSN 1873-0604.
\newblock \doi{https://doi.org/10.1002/nsg.12146}.
\newblock URL \url{https://www.earthdoc.org/content/journals/10.1002/nsg.12146}.

\bibitem[Eppelbaum(2010)]{Eppelbaum(2010)}
Lev Eppelbaum.
\newblock Application of microgravity at archaeological sites in israel: Some estimation derived from 3‐d modeling and quantitative analysis of gravity field.
\newblock volume~1, 03 2010.
\newblock \doi{10.4133/1.3176721}.

\bibitem[Carbone et~al.(2020)Carbone, Antoni-Micollier, Hammond, de~Zeeuw~van Dalfsen, Rivalta, Bonadonna, Messina, Lautier-Gaud, Toland, Koymans, Anastasiou, Bramsiepe, Cannavò, Contrafatto, Frischknecht, Greco, Marocco, Middlemiss, Ménoret, Noack, Passarelli, Paul, Prasad, Siligato, and Vermeulen]{Cabone(2020)}
D.~Carbone, L.~Antoni-Micollier, G.~Hammond, E.~de~Zeeuw~van Dalfsen, E.~Rivalta, C.~Bonadonna, A.~Messina, J.~Lautier-Gaud, K.~Toland, M.~Koymans, K.~Anastasiou, S.~Bramsiepe, F.~Cannavò, D.~Contrafatto, C.~Frischknecht, F.~Greco, G.~Marocco, R.~Middlemiss, V.~Ménoret, A.~Noack, L.~Passarelli, D.~Paul, A.~Prasad, G.~Siligato, and P.~Vermeulen.
\newblock The newton-g gravity imager: Toward new paradigms for terrain gravimetry.
\newblock \emph{Frontiers in Earth Science}, 8, 2020.
\newblock ISSN 2296-6463.
\newblock \doi{10.3389/feart.2020.573396}.
\newblock URL \url{https://www.frontiersin.org/article/10.3389/feart.2020.573396}.

\bibitem[Gadirov et~al.(2022)Gadirov, Kalkan, Ozdemir, Palabiyik, and Gadirov]{Gadirov(2022)}
Vagif Gadirov, Ekrem Kalkan, Adil Ozdemir, Yildiray Palabiyik, and Kamran Gadirov.
\newblock Use of gravity and magnetic methods in oil and gas exploration: Case studies from azerbaijan.
\newblock 4:\penalty0 143--156, 07 2022.

\bibitem[Zhdanov(2002)]{Zhdanov(2002)}
Michael~S. Zhdanov.
\newblock Chapter 1 - forward and inverse problems in geophysics.
\newblock In \emph{Geophysical Inverse Theory and Regularization Problems}, volume~36 of \emph{Methods in Geochemistry and Geophysics}, pages 3--28. Elsevier, 2002.
\newblock \doi{https://doi.org/10.1016/S0076-6895(02)80038-5}.
\newblock URL \url{https://www.sciencedirect.com/science/article/pii/S0076689502800385}.

\bibitem[Sen and Stoffa(2013)]{Sen(2013)}
M.K. Sen and P.L. Stoffa.
\newblock \emph{Global Optimization Methods in Geophysical Inversion}.
\newblock Cambridge University Press, 2013.
\newblock ISBN 9781107011908.
\newblock URL \url{https://books.google.co.uk/books?id=GDazJPWr3K4C}.

\bibitem[Lines and Treitel(1984)]{Lines(1984)}
L.R. Lines and S.~Treitel.
\newblock A review of least-squares inversion and its application to geophysical problems.
\newblock \emph{Geophysical Prospecting}, 32\penalty0 (2):\penalty0 159--186, 1984.
\newblock \doi{https://doi.org/10.1111/j.1365-2478.1984.tb00726.x}.
\newblock URL \url{https://onlinelibrary.wiley.com/doi/abs/10.1111/j.1365-2478.1984.tb00726.x}.

\bibitem[Tikhonov and Arsenin(1977)]{Tikhonov(1977)}
A.N. Tikhonov and V.I.A. Arsenin.
\newblock \emph{Solutions of Ill-posed Problems}.
\newblock Halsted Press book. Winston, 1977.
\newblock ISBN 9780470991244.
\newblock URL \url{https://books.google.co.uk/books?id=ECrvAAAAMAAJ}.

\bibitem[Zhdanov(2009)]{Zhdanov(2009)}
Michael~S. Zhdanov.
\newblock New advances in regularized inversion of gravity and electromagnetic data.
\newblock \emph{Geophysical Prospecting}, 57\penalty0 (4):\penalty0 463--478, 2009.
\newblock \doi{https://doi.org/10.1111/j.1365-2478.2008.00763.x}.
\newblock URL \url{https://onlinelibrary.wiley.com/doi/abs/10.1111/j.1365-2478.2008.00763.x}.

\bibitem[Bayes(1763)]{Bayes(1763)}
T.~Bayes.
\newblock An essay towards solving a problem in the doctrine of chances.
\newblock \emph{Philosophical Transactions}, 53:\penalty0 370--418, 1763.

\bibitem[Sivia and Skilling(2006)]{Sivia(2006)}
D.~S. Sivia and J.~Skilling.
\newblock \emph{Data analysis: a Bayesian tutorial}.
\newblock Oxford University Press, Oxford, 2nd edition, 2006.

\bibitem[Brown(2015)]{Brown(2015)}
G~Brown.
\newblock Bayesian mass anomaly estimation with measurements of gravity.
\newblock In \emph{2nd IET International Conference on Intelligent Signal Processing 2015 (ISP)}, pages 1--6, 2015.
\newblock \doi{10.1049/cp.2015.1754}.

\bibitem[Rodgers(2017)]{Rodgers(2017)}
Anthony~David Rodgers.
\newblock \emph{Bayesian inference from terrestrial gravimetry measurements of near-surface anomalies using a bespoke reversible-jump Markov chain Monte Carlo algorithm}.
\newblock PhD thesis, University of Birmingham, 2017.

\bibitem[Rezende and Mohamed(2016)]{Rezende(2016)}
Danilo~Jimenez Rezende and Shakir Mohamed.
\newblock Variational inference with normalizing flows, 2016.

\bibitem[Valentine and Sambridge(2022)]{Valentine(2022)}
Andrew Valentine and Malcolm Sambridge.
\newblock Emerging directions in geophysical inversion, 2022.

\bibitem[Yang et~al.(2022)Yang, Hu, Liu, Jie, Wang, and Chen]{Yang(2022)}
Qianguo Yang, Xiangyun Hu, Shuang Liu, Qu~Jie, Huaijiang Wang, and Qiuhua Chen.
\newblock 3-d gravity inversion based on deep convolution neural networks.
\newblock \emph{IEEE Geoscience and Remote Sensing Letters}, 19:\penalty0 1--5, 2022.
\newblock \doi{10.1109/LGRS.2020.3047131}.

\bibitem[Zhang et~al.(2022)Zhang, Zhang, Liu, and Fan]{Zhang(2022)}
Lianzhi Zhang, Guibin Zhang, Yao Liu, and Zhenyu Fan.
\newblock Deep learning for 3-d inversion of gravity data.
\newblock \emph{IEEE Transactions on Geoscience and Remote Sensing}, 60:\penalty0 1--18, 2022.
\newblock \doi{10.1109/TGRS.2021.3110606}.

\bibitem[Zhao et~al.(2021)Zhao, Curtis, and Zhang]{Zhao(2021)}
Xuebin Zhao, Andrew Curtis, and Xin Zhang.
\newblock {Bayesian seismic tomography using normalizing flows}.
\newblock \emph{Geophysical Journal International}, 228\penalty0 (1):\penalty0 213--239, 07 2021.
\newblock ISSN 0956-540X.
\newblock \doi{10.1093/gji/ggab298}.
\newblock URL \url{https://doi.org/10.1093/gji/ggab298}.

\bibitem[Hastings(1970)]{Hastings(1970)}
W.~K. Hastings.
\newblock {Monte Carlo sampling methods using Markov chains and their applications}.
\newblock \emph{Biometrika}, 57\penalty0 (1):\penalty0 97--109, 04 1970.
\newblock ISSN 0006-3444.
\newblock \doi{10.1093/biomet/57.1.97}.
\newblock URL \url{https://doi.org/10.1093/biomet/57.1.97}.

\bibitem[Skilling(2006)]{Skilling(2006)}
John Skilling.
\newblock Skilling, j.: Nested sampling for general bayesian computation. bayesian anal. 1(4), 833-860.
\newblock \emph{Bayesian Analysis}, 1:\penalty0 833--860, 12 2006.
\newblock \doi{10.1214/06-BA127}.

\bibitem[Buchner(2023)]{Buchner(2023)}
Johannes Buchner.
\newblock Nested sampling methods, 2023.

\bibitem[Ghalenoei et~al.(2021)Ghalenoei, Dettmer, Ali, and Kim]{Ghalenoei(2021)}
Emad Ghalenoei, Jan Dettmer, Mohammed~Y Ali, and Jeong~Woo Kim.
\newblock {Gravity and magnetic joint inversion for basement and salt structures with the reversible-jump algorithm}.
\newblock \emph{Geophysical Journal International}, 227\penalty0 (2):\penalty0 746--758, 07 2021.
\newblock ISSN 0956-540X.
\newblock \doi{10.1093/gji/ggab251}.
\newblock URL \url{https://doi.org/10.1093/gji/ggab251}.

\bibitem[Papamakarios et~al.(2021)Papamakarios, Nalisnick, Rezende, Mohamed, and Lakshminarayanan]{Papamakarios(2021)}
George Papamakarios, Eric Nalisnick, Danilo~Jimenez Rezende, Shakir Mohamed, and Balaji Lakshminarayanan.
\newblock Normalizing flows for probabilistic modeling and inference, 2021.

\bibitem[Li and Chouteau(1998)]{Li(1998)}
Xiong Li and Michel Chouteau.
\newblock Three-dimensional gravity modeling in all space.
\newblock \emph{Surveys in Geophysics}, 19, 1998.
\newblock \doi{10.1023/A:1006554408567}.
\newblock URL \url{https://doi.org/10.1023/A:1006554408567}.

\bibitem[Ashton et~al.(2019)Ashton, Hübner, Lasky, Talbot, Ackley, Biscoveanu, Chu, Divakarla, Easter, Goncharov, Vivanco, Harms, Lower, Meadors, Melchor, Payne, Pitkin, Powell, Sarin, Smith, and Thrane]{bilby}
Gregory Ashton, Moritz Hübner, Paul~D. Lasky, Colm Talbot, Kendall Ackley, Sylvia Biscoveanu, Qi~Chu, Atul Divakarla, Paul~J. Easter, Boris Goncharov, Francisco~Hernandez Vivanco, Jan Harms, Marcus~E. Lower, Grant~D. Meadors, Denyz Melchor, Ethan Payne, Matthew~D. Pitkin, Jade Powell, Nikhil Sarin, Rory J.~E. Smith, and Eric Thrane.
\newblock Bilby: A user-friendly bayesian inference library for gravitational-wave astronomy.
\newblock \emph{The Astrophysical Journal Supplement Series}, 241\penalty0 (2):\penalty0 27, apr 2019.
\newblock \doi{10.3847/1538-4365/ab06fc}.
\newblock URL \url{https://doi.org/10.3847%2F1538-4365%2Fab06fc}.

\bibitem[Koposov et~al.(2023)Koposov, Speagle, Barbary, Ashton, Bennett, Buchner, Scheffler, Cook, Talbot, Guillochon, Cubillos, Ramos, Johnson, Lang, Ilya, Dartiailh, Nitz, McCluskey, Archibald, Deil, Foreman-Mackey, Goldstein, Tollerud, Leja, Kirk, Pitkin, Sheehan, Cargile, Patel, and Angus]{dynesty}
Sergey Koposov, Josh Speagle, Kyle Barbary, Gregory Ashton, Ed~Bennett, Johannes Buchner, Carl Scheffler, Ben Cook, Colm Talbot, James Guillochon, Patricio Cubillos, Andrés~Asensio Ramos, Ben Johnson, Dustin Lang, Ilya, Matthieu Dartiailh, Alex Nitz, Andrew McCluskey, Anne Archibald, Christoph Deil, Dan Foreman-Mackey, Danny Goldstein, Erik Tollerud, Joel Leja, Matthew Kirk, Matt Pitkin, Patrick Sheehan, Phillip Cargile, Ruskin Patel, and Ruth Angus.
\newblock joshspeagle/dynesty: v2.1.2, June 2023.
\newblock URL \url{https://doi.org/10.5281/zenodo.7995596}.

\bibitem[Williams et~al.(2023)Williams, jmcginn, federicostak, and Veitch]{glasflow}
Michael~J. Williams, jmcginn, federicostak, and John Veitch.
\newblock uofgravity/glasflow: v0.2.0, February 2023.
\newblock URL \url{https://doi.org/10.5281/zenodo.7598678}.

\bibitem[Paszke et~al.(2019)Paszke, Gross, Massa, Lerer, Bradbury, Chanan, Killeen, Lin, Gimelshein, Antiga, Desmaison, Kopf, Yang, DeVito, Raison, Tejani, Chilamkurthy, Steiner, Fang, Bai, and Chintala]{pytorch}
Adam Paszke, Sam Gross, Francisco Massa, Adam Lerer, James Bradbury, Gregory Chanan, Trevor Killeen, Zeming Lin, Natalia Gimelshein, Luca Antiga, Alban Desmaison, Andreas Kopf, Edward Yang, Zachary DeVito, Martin Raison, Alykhan Tejani, Sasank Chilamkurthy, Benoit Steiner, Lu~Fang, Junjie Bai, and Soumith Chintala.
\newblock Pytorch: An imperative style, high-performance deep learning library.
\newblock In \emph{Advances in Neural Information Processing Systems 32}, pages 8024--8035. Curran Associates, Inc., 2019.
\newblock URL \url{http://papers.neurips.cc/paper/9015-pytorch-an-imperative-style-high-performance-deep-learning-library.pdf}.

\bibitem[Durkan et~al.(2020)Durkan, Bekasov, Murray, and Papamakarios]{nflows}
Conor Durkan, Artur Bekasov, Iain Murray, and George Papamakarios.
\newblock {nflows}: normalizing flows in {PyTorch}, November 2020.
\newblock URL \url{https://doi.org/10.5281/zenodo.4296287}.

\bibitem[Dinh et~al.(2017)Dinh, Sohl-Dickstein, and Bengio]{Dinh(2017)}
Laurent Dinh, Jascha Sohl-Dickstein, and Samy Bengio.
\newblock Density estimation using real nvp, 2017.

\bibitem[Biewald(2020)]{wandb}
Lukas Biewald.
\newblock Experiment tracking with weights and biases, 2020.
\newblock URL \url{https://www.wandb.com/}.
\newblock Software available from wandb.com.

\bibitem[Rakoczi(2023)]{flowinv}
Henrietta Rakoczi.
\newblock {G.I.Flow}, 11 2023.
\newblock URL \url{https://github.com/rakoczi-h/flowinv}.

\bibitem[Samantha R~Cook and Rubin(2006)]{Cook(2006)}
Andrew~Gelman Samantha R~Cook and Donald~B Rubin.
\newblock Validation of software for bayesian models using posterior quantiles.
\newblock \emph{Journal of Computational and Graphical Statistics}, 15\penalty0 (3):\penalty0 675--692, 2006.
\newblock \doi{10.1198/106186006X136976}.
\newblock URL \url{https://doi.org/10.1198/106186006X136976}.

\bibitem[Talts et~al.(2020)Talts, Betancourt, Simpson, Vehtari, and Gelman]{Talts(2020)}
Sean Talts, Michael Betancourt, Daniel Simpson, Aki Vehtari, and Andrew Gelman.
\newblock Validating bayesian inference algorithms with simulation-based calibration, 2020.

\bibitem[Dinh et~al.(2020)Dinh, Sohl-Dickstein, Larochelle, and Pascanu]{Dinh(2020)}
Laurent Dinh, Jascha Sohl-Dickstein, Hugo Larochelle, and Razvan Pascanu.
\newblock A rad approach to deep mixture models, 2020.

\end{thebibliography}






\end{document}